\documentclass[a4paper,12pt]{article}
\usepackage[utf8]{inputenc}
\usepackage{amsmath,amssymb}
\usepackage{graphicx}
\usepackage{float}
\usepackage{xcolor}
\usepackage[colorlinks=true, linkcolor=magenta, citecolor=red, urlcolor=black]{hyperref}
\usepackage{fancyhdr}
\allowdisplaybreaks

\begin{document}

\begin{center}
    {\LARGE \bfseries Absence of antisymmetric tensor fields: Clue from $f(R)$ model of gravity \par}
    \vskip 1em
    {\large
    Sonej Alam$^{1}$\thanks{e-mail: sonejalam36@gmail.com},
    Somasri Sen$^{1}$\thanks{e-mail: ssen@jmi.ac.in},
    Soumitra Sengupta$^{2}$\thanks{e-mail: tpssg@iacs.res.in}
    \par}
    \vskip 1em
    {\small
    $^{1}$Department of Physics, Jamia Millia Islamia, New Delhi 110025, India \\
    $^{2}$School of Physical Sciences, Indian Association for the Cultivation of Science, Kolkata 700032, India
    }
\end{center}

\vskip 1em

% ----- Footer (optional, can remove if you only want emails above) -----
\thispagestyle{fancy}
\fancyhf{} % clear header/footer
\fancyfoot[C]{\footnotesize
\begin{flushleft}
\textit{Email:}\\
$^{1}$\href{mailto:sonejalam36@gmail.com}{sonejalam36@gmail.com},\\
$^{2}$\href{mailto:ssen@jmi.ac.in}{ssen@jmi.ac.in},\\
$^{3}$\href{mailto:tpssg@iacs.res.in}{tpssg@iacs.res.in}
\end{flushleft}}

\renewcommand{\headrulewidth}{0pt}
\renewcommand{\footrulewidth}{0pt}

% ------------------- Your content -------------------
\noindent

%\maketitle

\begin{abstract}
One of the surprising aspects of the present Universe, is the absence of any noticeable observable effects of higher-rank antisymmetric tensor fields in any natural phenomena. Here, we address the possible explanation  of the absence  of the higher rank antisymmetric tensor fields within  the framework of a general class of $f(R)$ gravity represented by $f (R) = R +\alpha_n R^n$. We explore the setup in Einstein frame, where the higher curvature is manifested in terms of a scalar field with a potential through a conformal transformation. The evolution of different cosmological parameters is studied in the background of FRW universe. We show that while different cosmological parameters mimic their standard behaviour at different epochs for different forms of higher curvature gravity (i.e. different values of n ), the positive values of the scalar field in the models provide an additional suppression for the massless modes of higher rank antisymmetric field, 
%which gives a natural justification for the invisibility of their signatures. 
The Starobinsky model identified with $n=2$, provides heavier suppression compared to the others $n\neq2$. The result does not change even with the inclusion of the Cosmological Constant. Thus, our result reveals that a general class of  modified gravity models can successfully explain the suppression of the massless modes of higher rank antisymmetric tensor fields leading to their invisibility in the present universe.
\end{abstract}

\section{Introduction}
We haven't observed any overt indications of antisymmetric tensor fields having an impact on anything that occurs naturally, which is a rather startling aspect of our Universe. In this context, the second-rank antisymmetric tensor field, known as the Kalb-Ramond (KR) field \cite{PhysRevD.9.2273}, has been studied extensively. Such field appears naturally to cancel gauge anomaly in superstring theory\cite{Polchinski_1998,Green_2012}. It has also been shown that the KR field has a natural geometric interpretation as space-time torsion in background geometry through an antisymmetric extension of the affine connection\cite{RevModPhys.48.393,doi:10.1142/2358,1999CQGra..16L..89M}. While it is simple to argue that the coupling of the KR field to matter should be {$1/M_p$} in Einstein gravity, which is the same as the coupling of gravity with matter, based on dimensional considerations, there hasn't been any experimental signature  of the KR field in  the observable Universe. As a result, it can be inferred that if such tensor fields exist, they must be greatly suppressed at the current scale of the Universe.

It was shown earlier that the presence of warped  extra spatial dimension may lead to suppression of antisymmetric tensor field of various ranks\cite{121101,121501,107901,31,124008}. While any extra-dimensional theory brings in a plethora of other predictions hitherto unobserved,  within the domain of 3+ 1 dimensional Einstein gravity , it is hard to explain the reason of the disappearance of the massless KR field in our observable universe. As an alternative, in this work, we want to explore a dynamical mechanism of this suppression of the KR field as well as the other higher rank anti-symmetric fields in the light of higher curvature $f(R)$ theory. The presence of suitable higher curvature terms in the gravity action is permissible as long as they satisfy diffeomorphic invariance \cite{RevModPhys.82.451,2010LRR....13....3D,PhR_2017,2007_IJGMM,RevModPhys.72.452}. Higher curvature and extended gravity theories have attracted considerable attention in addressing various cosmological and astrophysical problems beyond standard General Relativity \cite{2016CQGra..33g5012P,2018EPJC...78..108D,2017arXiv170605964B,2016AnPhy.373...96B,2016EPJC...76..552C,2015PhRvD..92b6008A,2008PhRvD..77d6009C,2017EPJC...77..862O,2009PhRvD..80l3528L,2010JCAP...08..021B,PhLB_2006,PhRvD_2011,Alruwaili:2025gbw,Alruwaili:2025bwf,Alruwaili:2025mrc,Alruwaili:2025daq}.In particular, $f(R)$ gravity has been extensively studied due to its capability to explain early universe inflation, late-time cosmic acceleration, and other cosmological phenomena. Higher order curvature terms in the gravity actions in such models are automatically suppressed by Planck scale which explains why at low energy only Einstein's term dominates over the others. However, at high energy or at the epoch of early universe , the higher curvature term will have significant presence and in this context Starobinsky model \cite{1980PhLB...91...99S} has been studied extensively in different context\cite{Starobinsky:1981vz,1979ZhPmR..30..719S,1987,2001,2019PhRvD,JCAP_2019,PhysRevD.98.104002}. In this work, we explore whether the dynamical evolution of the components in the presence of higher curvature gravity in 3+1 dimension may lead to a natural geometric suppression of the massless higher rank antisymmetric tensor fields.

In section 2, we present the standard formalism of transformation of an action in $f(R)$ gravity to Einstein gravity through a conformal transformation which manifests the higher curvature in terms of a scalar degree of freedom along with a potential. This transformation has been extensively discussed in the literature\cite{RevModPhys.82.451,2010LRR....13....3D,2016CQGra..33g5012P}. In section 3, we present the field equations of the transformed action in an isotropic and homogeneous universe. The conformal transformation induces coupling of scalar field  with different component. The coupled nonlinear equations are presented as a set of autonomous system in terms of a new set of dynamical variables. The form of $f(R)$ and the potentials generated from this modified gravity are  discussed in section 4. Section 5 presents the solutions of the field equations in three different cases - A) $n=2$ or the Starobinsky model, B) $n\neq 2$ or generalised Starobinsky model and C) Both the models with Cosmological constant. In section 6 we highlight the finding of our work and finally, we discuss and conclude in section 7.

\section{$f(R)$ Formalism and Conformal Transformation}
For a rank $n$ antisymmetric tensor field $X_{a_1a_2...a_n}$, the field strength tensor of the massless mode can be written as $Y_{a_1a_2...a_{n+1}}=\partial_{[a_{n+1}}X_{a_1a_2...a_n]}$. In 3+1 dimensional spacetime, the maximum rank of an antisymmtric tensor field can be 3 as the field strength vanishes beyond that. In an early universe  post reheating epoch (around the time of nucleosynthesis), the effective action of higher curvature gravity in 3+1 dimension along with other fields can be written as

\begin{equation}
\begin{split}
S_{\text{gravity}} = \int d^4x \sqrt{-g} \bigg[ 
&\frac{F(R)}{2\kappa^2} + \bar{\Psi} i \gamma^\mu D_\mu \Psi - \frac{1}{4} F_{\mu\nu}F^{\mu\nu} \\[4pt]
& - \frac{1}{4} H_{\mu\nu\rho} H^{\mu\nu\rho}  - \frac{1}{4} T_{\mu\nu\rho\delta} T^{\mu\nu\rho\delta} 
\bigg]
\end{split}
\label{eq:action1}
\end{equation}

%\end{widetext}

\noindent
where the last 4 terms are the kinetic lagrangian of spin $\frac{1}{2}$ fermion fields $\psi$, U(1) gauge field $A_\mu$, Kalb Ramond field $B_{\mu\nu}$ and rank 3 antisymmetric tensor field $X_{\mu\nu\rho}$ respectively. $H_{\mu\nu\rho}$ and $T_{\mu\nu\rho\delta}$ are the field strength tensor of the Kalb Ramond field and rank 3 antisymmetric tensor field respectively defined as 
$H_{\mu\nu\rho}=\partial_{[\mu} B_{\nu\rho]}$  and $T_{\mu\nu\rho\delta}=\partial_{[\mu} X_{\nu\rho\delta]}$. \\

It is standard to represent \cite{RevModPhys.82.451,2010LRR....13....3D} the action (\ref{eq:action1}) in the following way with an introduction of an auxiliary field
\begin{equation}
\begin{split}
S = \int d^4x \sqrt{-g} \Bigg[ 
&\frac{1}{2\kappa^2} \phi R - V(\phi) + \bar{\Psi} i \gamma^\mu D_\mu \Psi - \frac{1}{4} F_{\mu\nu} F^{\mu\nu} \\[4pt]
& - \frac{1}{4} H_{\mu\nu\rho} H^{\mu\nu\rho} - \frac{1}{4} T_{\mu\nu\rho\delta} T^{\mu\nu\rho\delta} 
\Bigg]
\end{split}
\label{eq:action3}
\end{equation}

\noindent
where
\begin{equation}
\begin{aligned}
    \phi\equiv f^\prime (R)~\text{and}~ V(\phi) = \frac{1}{2\kappa^2}[{\phi~R - f(R)}]
\end{aligned}
\label{eq:pot1}
\end{equation}
% we can rewrite equation (\ref{eq:action2}) as

Now,  following a conformal transformation of the form 
$\tilde{g}_{\mu\nu}(x) = \Omega^2 {g}_{\mu\nu}(x)$, where  $\Omega^2=f'=\phi$ one can rewrite the action (\ref{eq:action3}) in the Einstein frame as \cite{PhysRevD.98.104002}
\begin{equation}
\begin{aligned}
S_E = & \int d^4x \sqrt{-\tilde{g}} \bigg( \frac{\tilde{R}}{2\kappa^2} 
- \frac{1}{2} \tilde{g}^{\mu\nu}\partial_\mu {\tilde\phi} \partial_\nu {\tilde\phi} 
- V(\tilde{\phi}) +e^{\sqrt{\frac{2}{3}} k \tilde\phi} \Psi^{+} \tilde{\gamma}^{0} i \tilde{\gamma}^\mu  D_{\mu} {\Psi}  - \frac{1}{4}\widetilde{g}^{\mu\alpha} \widetilde{g}^{\nu\beta} { F}_{\mu\nu}{F}_{\alpha\beta}  \\
& - \frac{1}{4}e^{-\sqrt{\frac{2}{3}} k \tilde\phi} \widetilde{g}^{\mu\alpha} \widetilde{g}^{\nu\beta} \widetilde{g}^{\rho\delta}{ H}_{\mu\nu\rho} {H}_{\alpha\beta \delta}   
 - \frac{1}{4}e^{-2\sqrt{\frac{2}{3}} k \tilde\phi} \widetilde{g}^{\mu\alpha} \widetilde{g}^{\nu\beta} \widetilde{g}^{\rho\delta}\widetilde{g}^{\sigma\lambda}{T}_{\mu\nu\rho\sigma} {T}_{\alpha\beta \delta\lambda} \bigg)\\
\end{aligned}
\label{eq:action4}
\end{equation}
\noindent
where the scalar field $\tilde\phi$ is defined as $\kappa\tilde\phi \equiv \sqrt{\frac{3}{2}} \ln \phi$ and  the metric in Einstein frame is represented by an overhead $\sim$. This choice of conformal transformation is consistent with $f'>0$. 

It should be noted that in the last two terms, which are the  field strength tensors of the massless mode of the rank 2 and 3 antisymmetric tensor fields, 
%{\color{red}apart from the usual \( 1/M_p \) %\cite{PhysRevD.98.104002} coupling an additional 
an exponential factor of $\tilde\phi$ %$e^{-\sqrt{\frac{2}{3}} k \tilde\phi(x)}$ 
  is present. So, the possible signatures of these field crucially depends on the nature of the scalar field $\tilde\phi$ in these exponential factors %($e^{-\sqrt{\frac{2}{3}} k \tilde\phi(x)}$).

(i) If the scalar field $\tilde\phi$ is positive then the exponential factor provides an additional suppression to these terms making the signature of the anti symmetric fields invisible.

(ii) Otherwise if $\tilde\phi$ is negative, the exponential factor will enhance the overall coupling, thereby making the signature significant.

{\it Hence we study the evolution of the scalar field $\tilde\phi$ to understand the nature of the field which in turn will determine whether the imprints of the higher rank antisymmtric tensor fields will be visible or not.} 

\section{Field Equations in Einstein Frame}

Our goal is to explore how the scalar field evolves cosmologically in the Einstein frame in a background FRW space-time in order to understand whether the coupling of the higher rank antisymmetric fields are suppressed or enhanced as the universe evolves. And to do so, we further rewrite the entire action into two parts - one part consists of the scalar field ($\tilde\phi$) in Einstein gravity and the other part contains all other matter fields, namely, fermions, gauge fields, and higher rank antisymmetric fields which are coupled to the scalar field $\tilde\phi$ through the metric $g_{\mu\nu}$. Note that we do not consider any back reaction of these fields on the scalar field. 
\begin{equation}
    \begin{aligned}
        S_E & = \int d^4x \sqrt{-\tilde{g}} \bigg( \frac{\tilde{R}}{2\kappa^2} - \frac{1}{2} \tilde{g}^{\mu\nu}\partial_\mu {\tilde\phi} \partial_\nu {\tilde\phi} - V(\tilde{\phi})\bigg)\\
        & +\int d^4x  ~{\cal L}_M(e^{-\sqrt{\frac{2}{3}} k \tilde\phi(x)}\tilde{g}_{\mu\nu},{\psi},{ A}_\mu,{B}_{\mu\nu},{X}_{\mu\nu\rho})\\
    \end{aligned}
    \label{eq:action5}
\end{equation}

\noindent
${{\cal L}_{\tilde{\phi}}}=\frac{1}{2} \tilde{g}^{\mu\nu}\partial_\mu {\tilde\phi} \partial_\nu {\tilde\phi} - V(\tilde{\phi})$ is the Lagrangian density of the scalar field ${\tilde{\phi}}$ and ${{\cal L}_M}$ is the Lagrangian density of 
matter fields. The energy-momentum tensor of the scalar field and matter, respectively is defined as,
\begin{equation}
\begin{aligned}
{\tilde{T}^{\tilde{\phi}}}_{\mu\nu}= -\frac{2}{\sqrt{-\tilde{g}}}\frac{\partial(\sqrt{-\tilde g}{{\cal L}_{\tilde \phi}})}{\partial{\tilde g}^{\mu\nu}} ~~\text{and} ~~ {\tilde {T}}^M_{\mu\nu}= -\frac{2}{\sqrt{-\tilde g}}\frac{\partial{\cal L}_M}{\partial{\tilde g}^{\mu\nu}}
\label{eq:em1}
\end{aligned}
\end{equation}
\noindent
Variation of the action (\ref{eq:action5}) with respect to the field $\tilde\phi$ gives the dynamical equation for the scalar field  
\begin{equation}
\begin{aligned}
\square{\tilde{\phi}}- \frac{\partial V}{\partial{\tilde{\phi}}}+\frac{1}{\sqrt{-\tilde g}}\frac{\partial{\cal L}_M}{\partial{\tilde {\phi}}}=0
\label{eq:sf1}
\end{aligned}
\end{equation}
\noindent 
where ${\cal{L}}_M
%\equiv{\cal{L}}_M(g_{\mu\nu})
\equiv{\cal{L}}_M(e^{-\sqrt{\frac{2}{3}} k \tilde\phi(x)}\tilde{g}_{\mu\nu})$. Hence 
\begin{equation}
\begin{aligned}
\frac{\partial{\cal L}_M}{\partial{\tilde {\phi}}}=\frac{\partial{\cal L}_M}{\partial g^{\mu\nu}}\frac{\partial g^{\mu\nu}} {\partial{\tilde {\phi}}} ={\sqrt{-\tilde g}}\kappa Q {\tilde T}
%{\tilde{g}}^{\mu\nu} {\tilde {T}}^M_{\mu\nu}   
\label{eq:sf2}
\end{aligned}
\end{equation}

where $Q=-\frac{1}{\sqrt{6}}$ is the strength of the coupling between the field and matter, which is constant for any form of $f(R)$.
%\vspace{0.5cm}

\noindent
The matter in (\ref{eq:em1}) is considered in the form of perfect fluid $\tilde T^{\mu (M)}_{\nu} = \text{diag}(- \tilde\rho_M, \tilde P_M, \tilde P_M,\tilde P_M)$.
$\tilde{\rho}_M$ and $\tilde{P}_M$ are the energy density and pressure of matter. 

We consider the observable universe to
%In the observable universe, we consider the total energy density $\tilde\rho_M$ 
consist of radiation and matter (baryonic and CDM) i.e., $\tilde\rho_M=\tilde\rho_r+\tilde\rho_m$.

In a flat FRW background the field equations are given by
\begin{equation}
 \begin{aligned}
     3\tilde H^{2} = \kappa^{2}[\frac{1}{2} (\frac{d\tilde\phi}{ d\tilde t})^2+ V(\tilde\phi) + \tilde{\rho}_M]
     \label{eq:h2}
 \end{aligned}   
\end{equation}
\begin{equation}
    \begin{aligned}
        \frac{d^2\tilde\phi}{d\tilde{t}^2} + 3\tilde{H}\frac{d\tilde\phi}{d\tilde{t}} + \frac{\partial V(\tilde{\phi})}{\partial{\tilde{\phi}}} = -\kappa Q(\tilde{\rho}_M - 3\tilde{P}_M)
        \label{eq:phi1}
    \end{aligned}
\end{equation}
\noindent
%The conservation equation for matter is 
\begin{equation}
    \begin{aligned}
        \frac{d\tilde\rho_M}{d\tilde t}+ 3 \tilde H (\tilde \rho_M + \tilde P_M) = \kappa Q(\tilde{\rho}_M - 3\tilde{P}_M) \frac{d\tilde\phi}{d\tilde t}
        \label{eq:rhom1}
    \end{aligned}
\end{equation}
\noindent
where  $\tilde H \equiv \frac{1}{\tilde a} \frac{d\tilde a}{\tilde dt}$. 
%Denoting the energy density of the field $\tilde \rho_{\tilde\phi} = (\frac{1}{2} (\frac{d^2\tilde\phi}{\tilde dt^2})+ V(\tilde\phi))$ and 
%the pressure of the field $\tilde P_{\tilde\phi} = (\frac{1}{2} (\frac{d^2\tilde\phi}{\tilde dt^2})- V(\tilde\phi))$, 
One can rewrite equation (\ref{eq:phi1}) as
\begin{equation}
    \begin{aligned}
        \frac{d\tilde\rho_\phi}{d\tilde t}+ 3 \tilde H (\tilde \rho_{\tilde\phi} + \tilde P_{\tilde\phi})  = -\kappa Q(\tilde{\rho}_M - 3\tilde{P}_M) \frac{d\tilde\phi}{d\tilde t}
    \label{eq:phi2}
    \end{aligned}
\end{equation}
%Considering the matter to be pressureless 
where $\tilde \rho_{\tilde\phi} = \frac{1}{2}\left(\frac{d\tilde\phi}{d\tilde t}\right)^2 + V(\tilde\phi)$ is the energy density of the scalar field and 
$\tilde P_{\tilde\phi} = \frac{1}{2}\left(\frac{d\tilde\phi}{d\tilde t}\right)^2 - V(\tilde\phi)$ is the pressure of the scalar field.
Using the standard thermodynamic relation $\tilde{P}_M=\omega\tilde{\rho}_M$, 
one can easily find a relation between $\tilde{\rho}_M$ and $\tilde{\phi}$ from equation (\ref{eq:rhom1}) 
\begin{equation}
    \begin{aligned}
        \tilde{\rho}_M =  \tilde{\rho}_{M0} \tilde a^{-3(1+\omega)} \exp({\frac{\kappa}{\sqrt{6}}(1-3\omega)(\tilde\phi - \tilde\phi_0)})
    \end{aligned}
    \label{eq:rhom2}
\end{equation}
So the set of dynamical equations to be handled are 
%representing the to be handled for finding out how the field rolls are
\begin{equation}
    \begin{aligned}
     3\tilde H^{2} = \kappa^{2}\bigg[\frac{1}{2}{\tilde\phi^{\prime 2}} \tilde H^{2}+ V(\tilde\phi) + \tilde{\rho}_m+\tilde\rho_r \bigg]
     \end{aligned}
     \label{eq:h2n}
\end{equation}
\begin{equation}
    \begin{aligned}
      \tilde\rho_\phi^\prime+ 3 (\tilde \rho_{\tilde\phi} + \tilde P_{\tilde\phi})  = \frac{\kappa}{\sqrt{6}} (\tilde{\rho}_m)\tilde\phi^\prime
      \end{aligned}
    \label{eq:phi2n}  
\end{equation}
\begin{equation}
    \begin{aligned}
      \tilde\rho_m^\prime+ 3 \tilde \rho_m = -\frac{\kappa}{\sqrt{6}}\tilde{\rho}_m \tilde\phi^\prime
       \end{aligned}
       \label{eq:rhom2}
\end{equation}
\begin{equation}
    \begin{aligned}
        \tilde\rho_r^\prime + 4 \tilde \rho_r =0
    \end{aligned}
    \label{eq:rhor}
\end{equation}

where prime denotes the differentiation with respect to $N (=\ln{\tilde a})$. Note that since the radiation field is traceless the scalar field does not couple to radiation (\ref{eq:sf2}). 
%Note that, since $\tilde\phi$ interacts with matter $\tilde\rho_m$, equations (\ref{eq:phi2n}) and (\ref{eq:rhom2}) are coupled whereas radiation is separately conserved (\ref{eq:rhor}).
In order to solve these coupled nonlinear inhomogeneous equations, we introduce a set of dynamical variables \cite{Copeland:1997et,delaMacorra:1999ff,Scherrer:2007pu,Scherrer:2008be,Sen:2009yh,Gupta:2011kw,2012PhLB..713..140H,2019JHEP...06..070B,2012PhLB..718....5A}
\begin{equation}
    \begin{aligned}
        x = \frac{\kappa \tilde\phi\prime}{\sqrt{6}}, 
        y = \frac{\kappa\sqrt{V}}{\sqrt{3}\tilde{H}},
        \lambda = - \frac{V_{,\tilde\phi}}{kV},~
        m = \frac{\kappa^2 \tilde \rho_m}{3 \tilde H^2},~
        r = \frac{\kappa^2 \tilde \rho_r}{3 \tilde H^2}
        %l = \frac{\kappa^2 \tilde \rho_\Lambda}{3 \tilde H^2}
    \end{aligned}
    \label{eq:var}
\end{equation}
and rewrite the equations (\ref{eq:h2n}-\ref{eq:rhor}) in terms of these new variables forming a set of autonomous systems of first-order differential equations 
\begin{eqnarray}
\begin{aligned}
\frac{dx}{dN} &= \frac{x}{2}(3x^2+3-3y^2 +r)+\sqrt{\frac{3}{2}} \lambda y^2\\
&-3x +\frac{1}{2}(1-x^2-y^2-r)\\
\frac{dy}{dN} &= -y[\sqrt{\frac{3}{2}} x\lambda +\frac{1}{2}(3y^2-3-3x^2-r)]\\
\frac{d\lambda}{dN} &=\sqrt{6}x\lambda^2[1-\Gamma]\\
\frac{dr}{dN} &=-r[4+(-3x^2+3y^2-r-3)]\\
%\frac{dl}{dN} &=l[4x-(-3x^2+3y^2-r+3l-3)]
\end{aligned}    
\label{eq:new-set2}
\end{eqnarray}
where, $\Gamma = \frac{V V_{\tilde\phi \tilde\phi}}{V_{\tilde\phi}^2}$.
In terms of these new variables, equation (\ref{eq:h2n}) takes the form
$1 = x^2 + y^2 + m + r $. 

\section{General class of $f(R)$ : $f(R) = R + \alpha_n R^n$}

\begin{figure}[h!]
    \centering
    \includegraphics[height=4.0cm, width=0.34\linewidth]{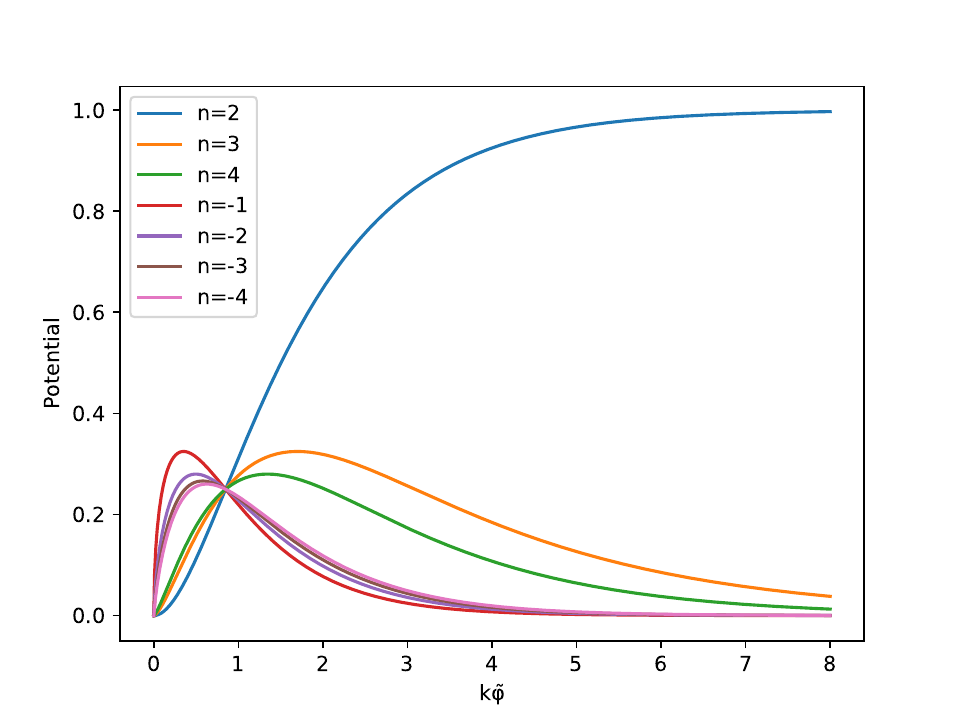}
    \includegraphics[height=4.0cm, width=0.32\linewidth]{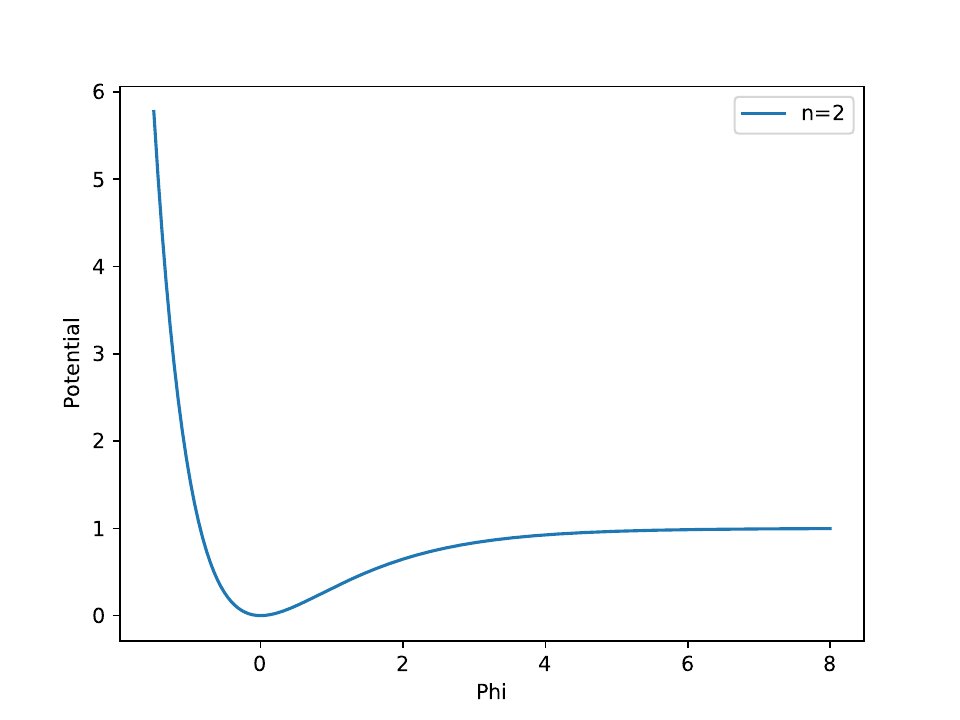}
    \includegraphics[height=4.0cm, width=0.32\linewidth]{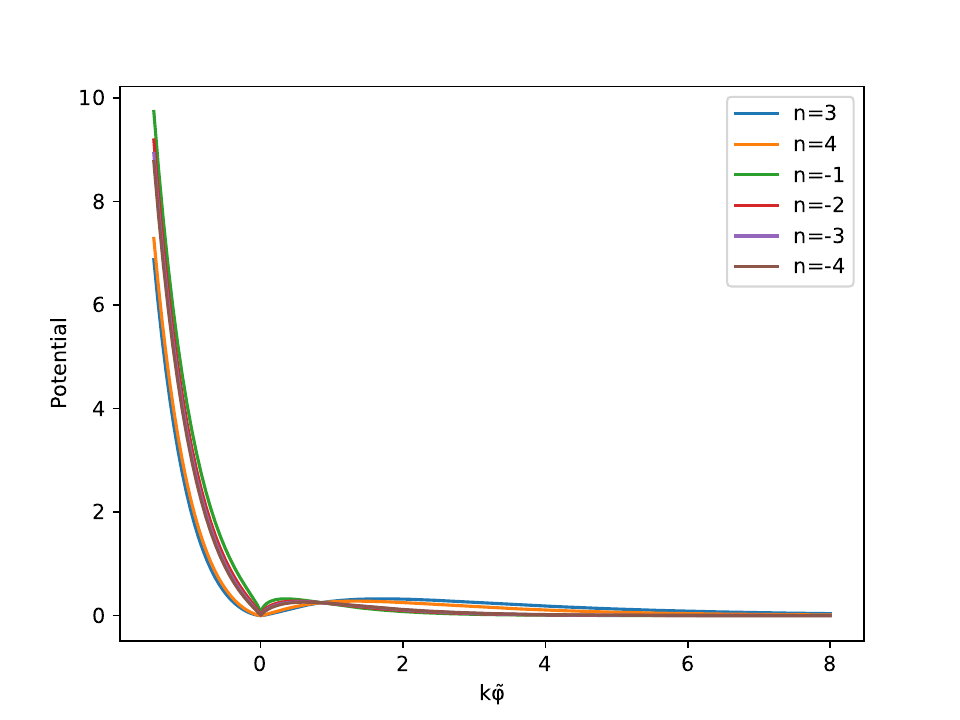}
    \caption{Form of potentials for different $n$}
    \label{fig:potn}
\end{figure}

In this work, we consider a very general class of modified gravity $f(R) = R + \alpha_n R^n$, where $\alpha_n$ has a suitable dimension. The Starobinsky model of modified gravity corresponds to $n=2$ i.e., $f(R) = R + \alpha_2 R^2$, where $\alpha_2 \sim \frac{1}{M^2}$, where $M$ is in MeV scale. With this general form of $f(R)$, the corresponding potential is
\begin{equation}
    \begin{aligned}
        V(\tilde\phi)={\cal A}e^{-2\sqrt{\frac{2}{3}}\kappa\tilde\phi}\bigg( e^{\sqrt{\frac{2}{3}}\kappa\tilde\phi}-1\bigg)^{\frac{n}{n-1}}
    \end{aligned}
    \label{eq:v}
\end{equation}
where ${\cal A}=\frac{(n-1)\alpha_n}{2\kappa^2}\big[\frac{1}{n\alpha_n}\big]^{\frac{n}{n-1}}$. The potential becomes zero 
 at $\tilde\phi=0$, irrespective of $n$. For $n=2$, the potential increases and becomes constant for positive $\tilde\phi$ . For negative $\tilde\phi$, it increases exponentially. However, for $n\neq2$ cases,  the potential increases initially, reaches a maxima but goes to zero as positive $\tilde\phi$ increases. In the left panel of fig (\ref{fig:potn}), we present the form of the potentials for various values of $n$ for positive $\tilde\phi$. In the center panel, we present the potential for $n=2$ for both positive and negative $\tilde\phi$. In the right panel we plotted the potential for the cases $n\neq 2$ for both positive and negative $\tilde\phi$.  In this figure, we have plotted the absolute value of the potential as it is not defined for negative  $\tilde\phi$. The form of the potentials are independent of the choice of $\cal A$.
 %the exponential term outside the bracket is the dominating term . For $n>2$, the power of term within the bracket, i.e, $(\frac{n}{n-1})$ is positive but less than 2 and for $n<0$ , it is positive but less than 1. So, 
% the potential increases sharply for negative $\tilde\phi$ and goes to $0$ for positive $\tilde\phi$. In fig (\ref{fig:potn})we present the form of potentials for various values of $n$. The form of the potentials are independent of the choice of $\cal A$.
%In order to find the cosmological evolution of $\tilde\phi$ with this potential one needs to work out with equations (\ref{eq:h2}) - (\ref{eq:phi2}). 
For the general potential $\Gamma$ takes a nontrivial form
\begin{equation}
    \Gamma = \frac{\frac{n}{(n-1)^2}-\frac{3n}{n-1}\left[1-e^{-\sqrt{\frac{2}{3}}\kappa\tilde\phi}\right]
    %-\frac{2n}{n-1}\left[1-e^{-\sqrt{\frac{2}{3}} \kappa\tilde\phi}\right]
    +4\left[1-e^{-\sqrt{\frac{2}{3}} \kappa\tilde\phi}\right]^2}{\left[\frac{n}{n-1}-2(1-e^{-\sqrt{\frac{2}{3}}\kappa \tilde \phi})\right]^2}
    \nonumber
\end{equation}

 We solve these equations numerically taking different integral values of $n$ ranging from -4 to 4
 with suitable initial conditions. Corresponding to different values of $n$, the expression for $\Gamma$ changes and hence the set of equations.

\section{Different cases of $n$}
{\bf \subsection{Case I: $n=2$, Starobinsky model}}

\begin{figure}[h!]
\centering
\includegraphics[height=4.0cm, width=0.32\linewidth]{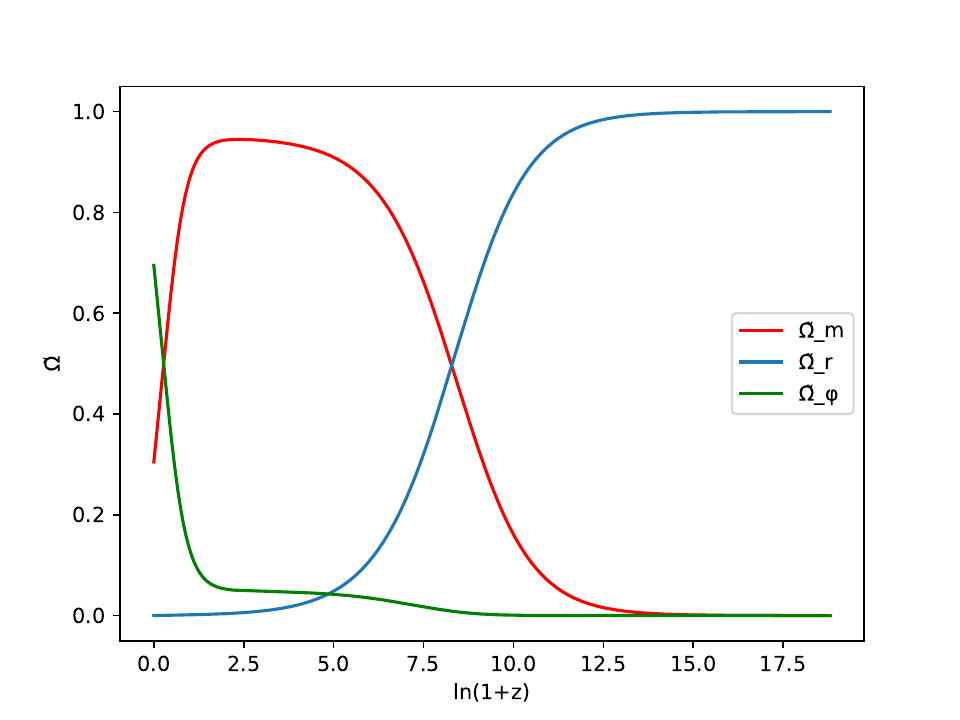}
\includegraphics[height=4.0cm, width=0.32\linewidth]{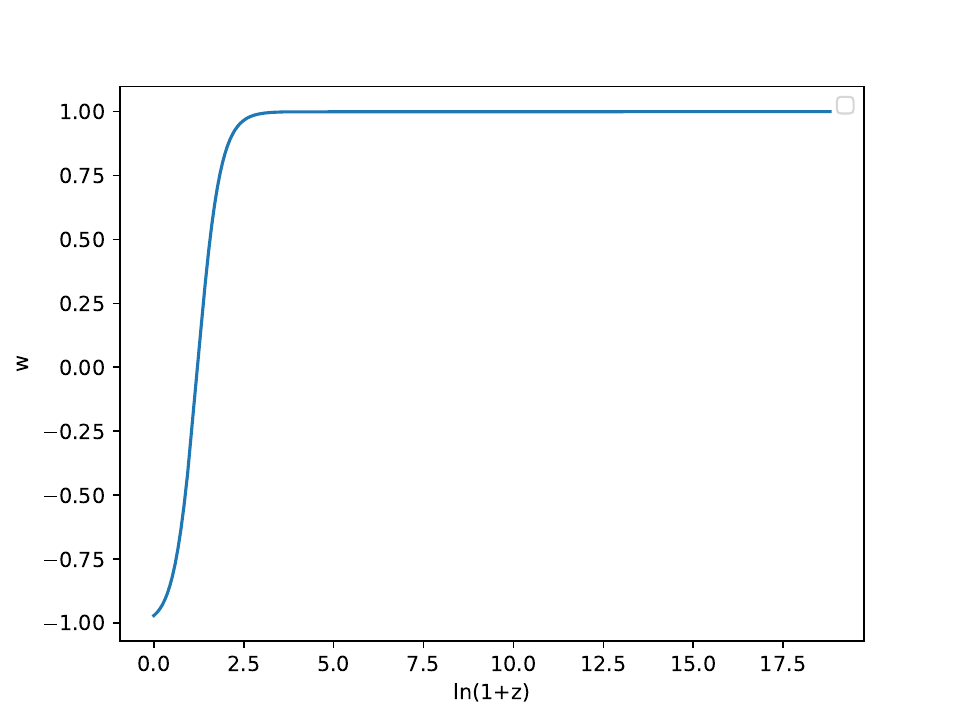}
\includegraphics[height=4.0cm, width=0.32\linewidth]{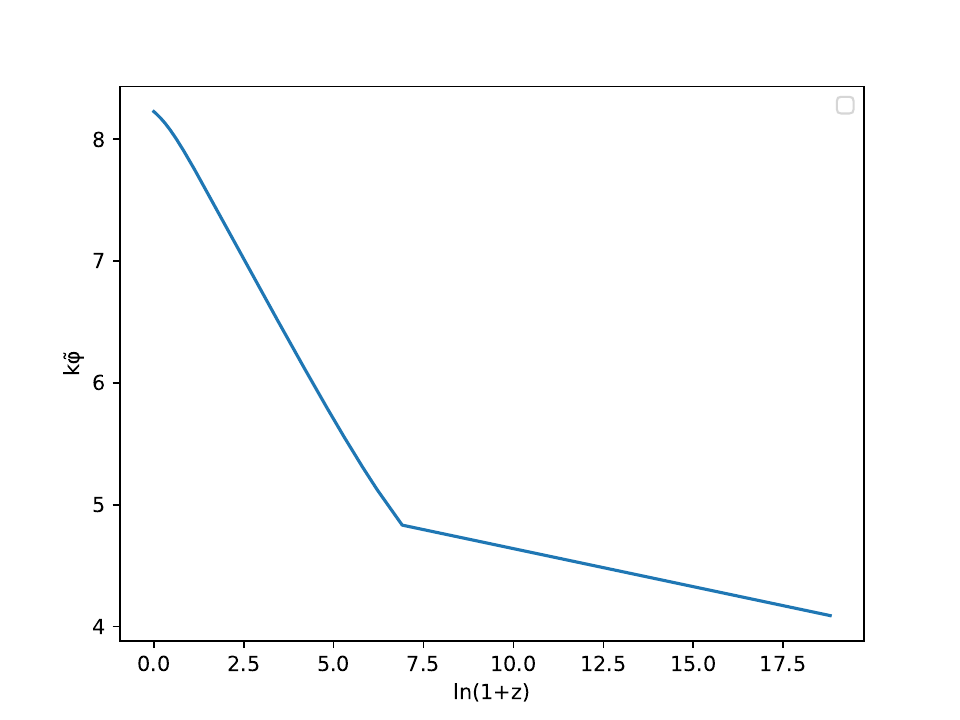}
\caption{Evolution of density parameters, scalar field $\tilde\phi$ and $\omega_{\tilde\phi}$ in Starobinsky model(n=2).  }
\label{fig:en2}
\end{figure}

For $n=2$, $f (R)$ takes the form $R +\alpha_{2} R^2$, which  is the famous Starobinsky model. Substituting the value of $n$, the $\frac{d\lambda}{dN}$ equation becomes
\begin{equation}
\begin{aligned}
%\frac{dx}{dN} &= \frac{x}{2}(3x^2-3-3y^2 +r - 3l)+\sqrt{\frac{3}{2}} \lambda y^2-\frac{1}{2}(1-x^2-y^2-r+3l)\\
%\frac{dy}{dN} &= -y[\sqrt{\frac{3}{2}} x\lambda +\frac{1}{2}(3y^2-3-3x^2-r+3l)]\\
\frac{d\lambda}{dN} =\sqrt{\frac{3}{2}}\lambda^2 x \left[1-\frac{1}{\lambda}\sqrt{\frac{8}{3}}\right]
%\frac{dr}{dN} &=-r[4+(-3x^2+3y^2-r+3l-3)]\\
%\frac{dl}{dN} &=l[4x-(-3x^2+3y^2-r+3l-3)]
\end{aligned}    
\label{eq:l-n1}
\end{equation}
Note that the other equations in (\ref{eq:new-set2}) remain the same. We solve the above equations numerically, with suitable initial conditions at the epoch of nucleosynthesis at $z\sim 10^8$. At such an early epoch radiation is the most dominating component compared to others . We have chosen \( \ r = 0.99997,\ x = 0.003,\ y = 2.2 \times 10^{-15}, \ \text{and} \ \lambda = -6 \times 10^{-2}\).
%These initial conditions are chosen such that radiation and matter components mimic the usual evolution in standard cosmology where 
Following these conditions, radiation-matter equality happens around redshift 3300 and matter in the present universe constitutes approximately 30\% of the energy density. In Fig (\ref{fig:en2}), we present different cosmological parameters and the scalar field as they vary with redshift. The left panel presents different density parameters with respect to $\ln(1+z)$, 
%which follows the behavior of standard cosmology. the evolution of the energy densities for all the components are given on the right. 
The density parameters of matter and radiation follow their behavior of standard cosmology, while the density parameter of $\tilde\phi$ field never dominates the evolution until recently. To further understand the behavior of $\tilde\phi$, in the middle panel we plot $\omega_{\tilde\phi}(=\frac{\tilde P_{\tilde\phi}}{\tilde\rho_{\tilde\phi}})$ with respect to $\ln(1+z)$. $\omega_{\tilde\phi}$ remains constant for a long period and starts varying late and finally approaches $-1$ at present. So, in the late time the field behaves as dark energy. This behaviour of $\omega_{\tilde\phi}$ resembles the equation of state of the thawing quintessence field. However, since our main goal is to study the nature of the scalar field, in the right most panel we present $\tilde\phi$ with respect to $\ln(1+z)$. We observe that $\tilde\phi$ remains positive along the entire evolution and does not vary much.
%Most of the time it scales as matter and becomes constant recently.

\begin{figure}[h!]
 \centering
\includegraphics[height=4.0cm, width=0.34\linewidth]{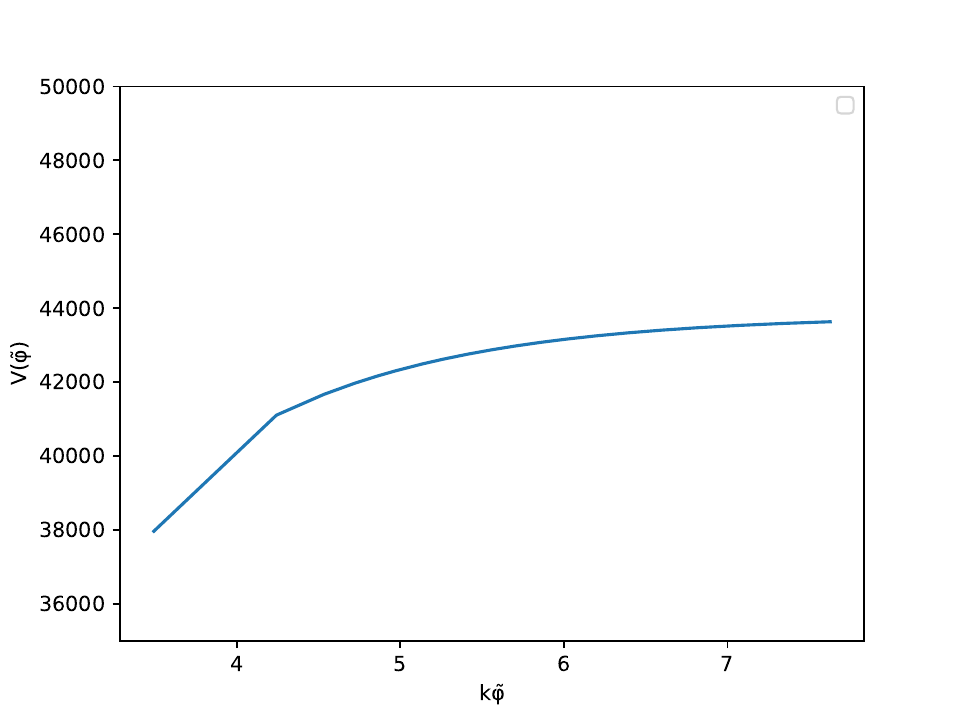}
\includegraphics[height=4.0cm, width=0.32\linewidth]{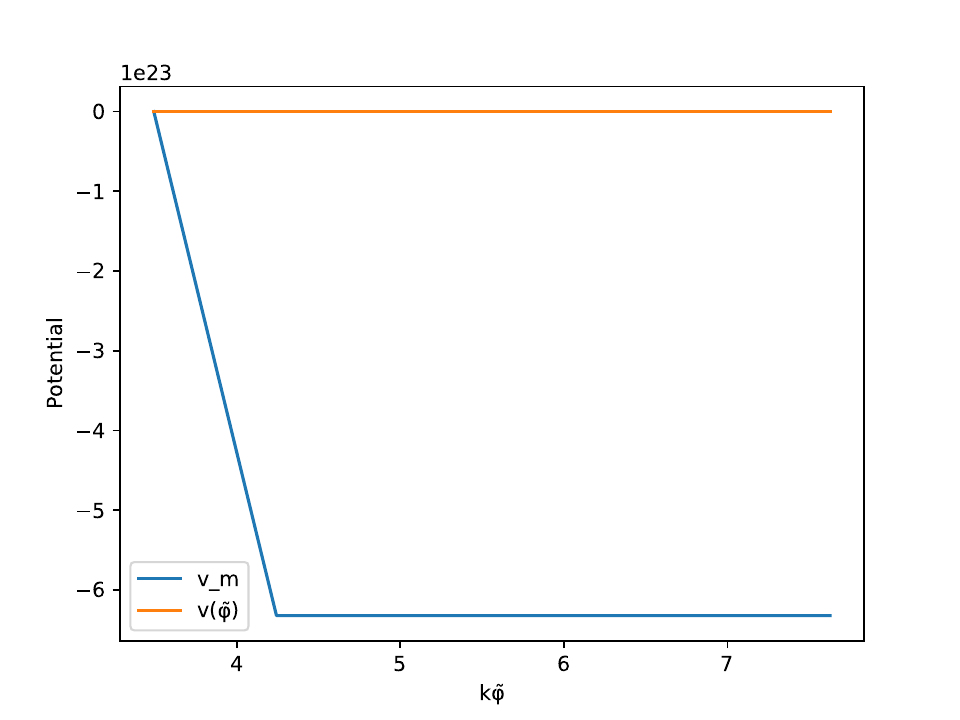}
\includegraphics[height=4.0cm, width=0.32\linewidth]{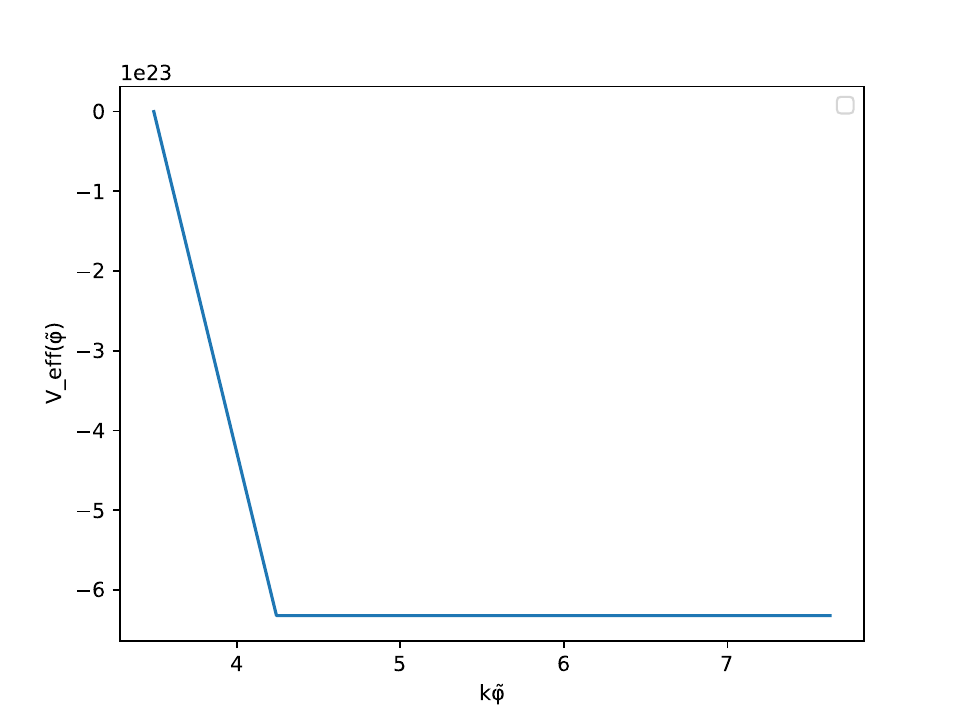}
\caption{Different potentials  $V(\tilde\phi)$, $V_m$ and $V_{eff}$ in Starobinsky model}
\label{fig:pot2}
 \end{figure}

To understand the behaviour of the scalar field better, we investigate the potential $V(\tilde\phi)$ and find that  the scalar field rolls up the potential $V(\tilde\phi)$ .
 This is due to the coupling between the scalar field and matter. Denoting the coupling term on the right hand side of equation (\ref{eq:phi1}) as $\frac{\partial V_m}{\partial\phi}=\kappa Q\tilde{\rho}_m$, one can write the potential $V_m$ due to the coupling term. Hence the scalar field rolls down the effective potential $V_{eff}(=V(\tilde\phi)+V_m$). Plotting both the terms $V(\tilde\phi)$ and $V_m$ separately, we realized that  $V_m$ is far more dominant than $V(\tilde\phi)$, the potential arising from modified gravity. Hence the scalar field rolls down the effective potential $V_{eff}(\approx V_m)$. In Fig (\ref{fig:pot2}), we present all the three different potentials $V(\tilde\phi)$, $V_m$ and $V_{eff}$.
 
 %However, $\tilde\phi$ remains positive along the entire evolution and does not vary much. Such values of $\tilde\phi$ in action (\ref{eq:action4}) ensure that the torsion is heavily suppressed, thereby failing to leave some imprints in the observable universe. 

{\bf  \subsection{Case II:  $n \neq 2$, $f (R) = R +\alpha_{n} R^n$}}
 
\begin{figure}[h!]
\centering
\includegraphics[height=4.0cm, width=0.34\linewidth]{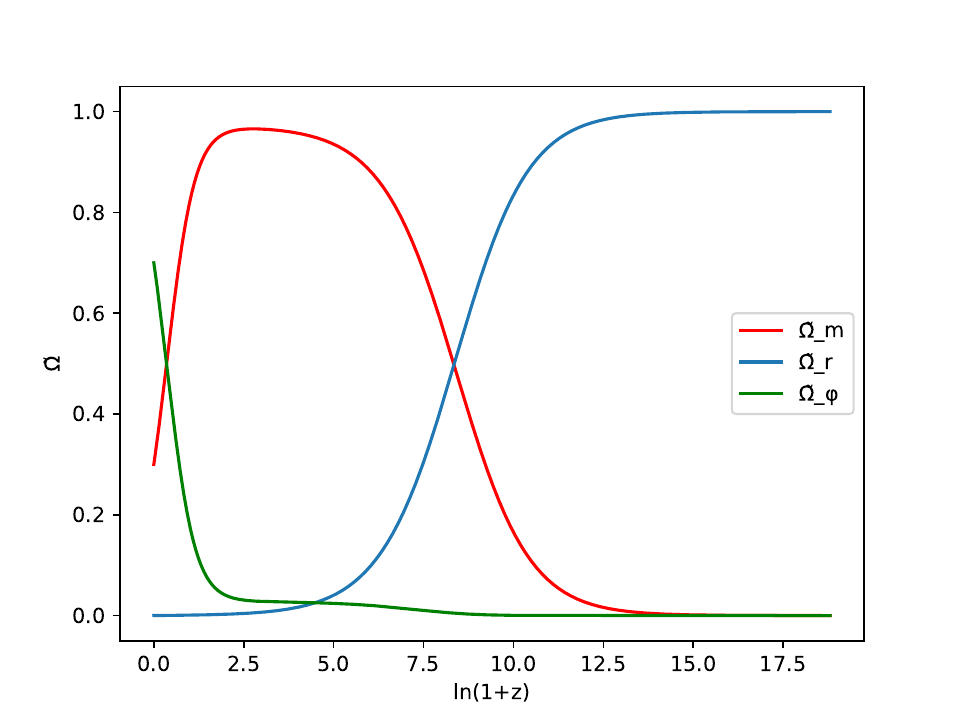}
\includegraphics[height=4.0cm, width=0.32\linewidth]{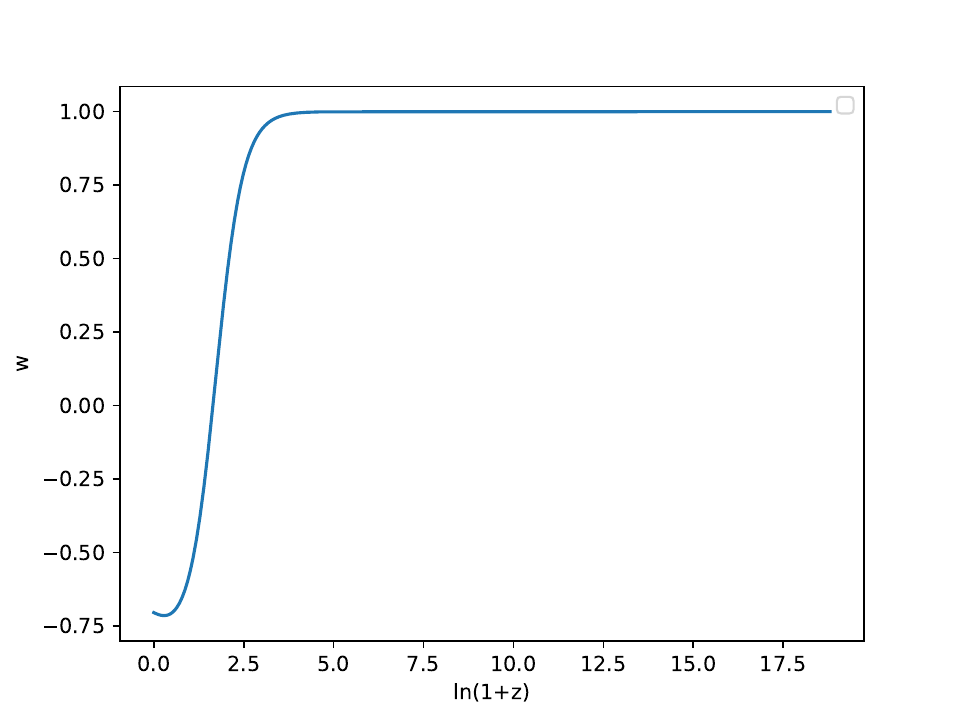}
\includegraphics[height=4.0cm, width=0.32\linewidth]{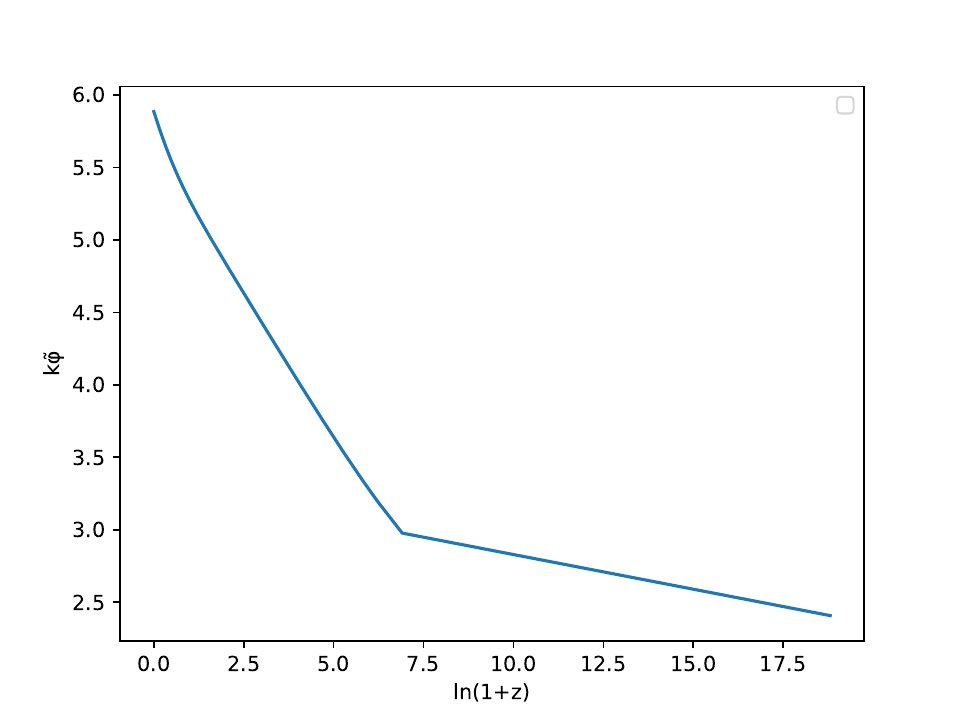}
\caption{Evolution of different cosmological parameters for $n=-2$ case}
\label{fig:n-2}
\end{figure}

\begin{figure}
\includegraphics[height=4.0cm, width=0.34\linewidth]{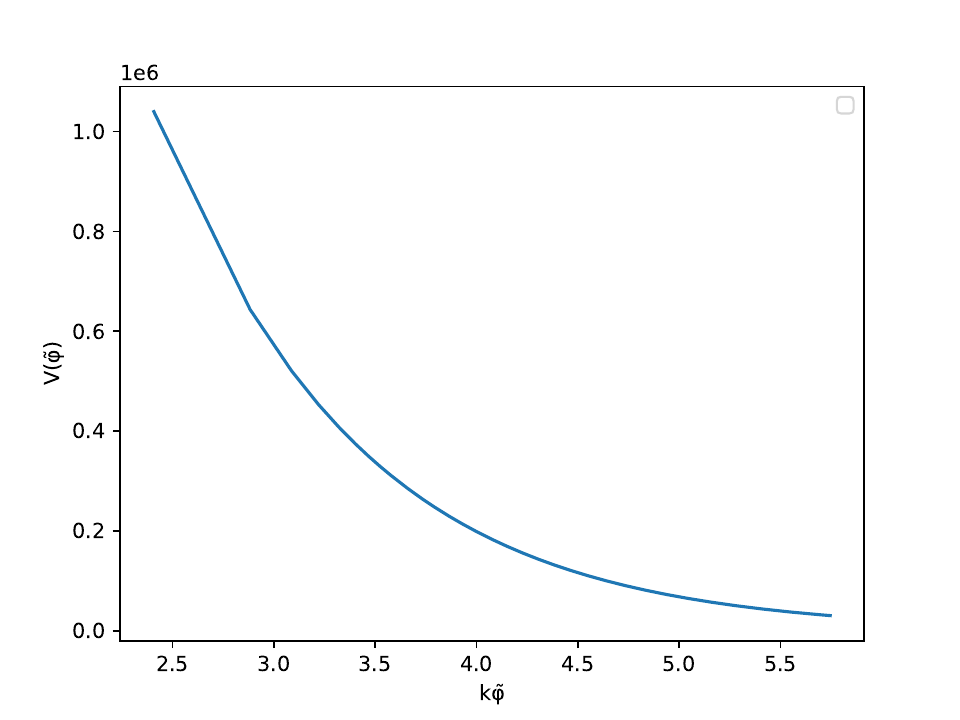}
\includegraphics[height=4.0cm, width=0.32\linewidth]{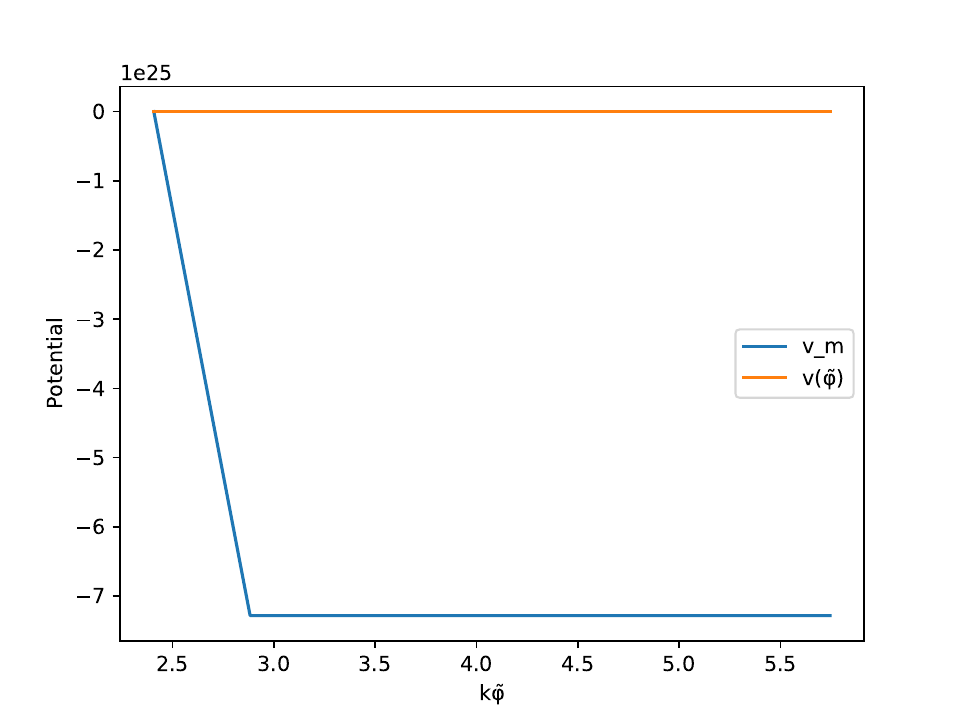}
\includegraphics[height=4.0cm, width=0.32\linewidth]{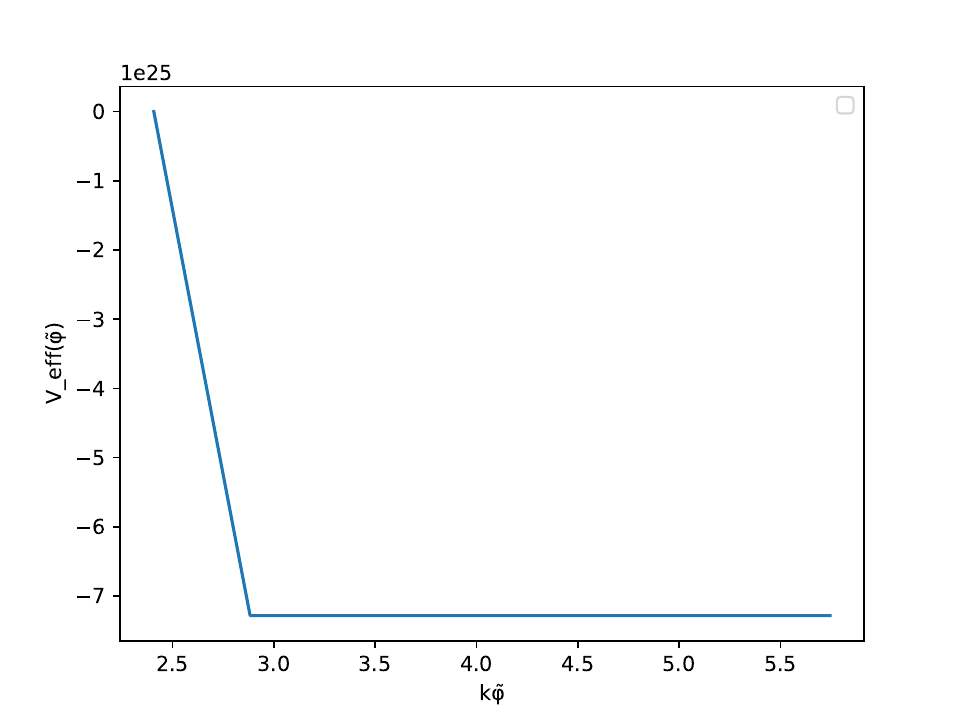}
\caption{Different potentials for $n=-2$ case}
\label{fig:n-2p}
 \end{figure}

 Similarly, we have studied the other cases where $n$ takes positive values $3,4$ as well as negative values $-1,-2,-3,-4$ in the modified gravity form $f(R) = R +\alpha_{n} R^n$ and suitably changed the set of dynamical equations. We have already noticed in fig (\ref{fig:potn}) that for all these cases, the forms of the potential are similar irrespective of different values of $n$. This hints that for all other cases $(n\neq2)$ the results are expected to be same. Here, we present only one case ($n=-2$) on behalf of the $n\neq 2$ cases. For $n=-2$, the $\frac{d\lambda}{dN}$ equation becomes
 \begin{equation}
     \begin{aligned}
\frac{d\lambda}{dN} =\sqrt{\frac{3}{8}}x(4\sqrt{8/3}-6\lambda)(\sqrt{8/3}-\lambda)
     \end{aligned}
     \label{eq:l-n2}
 \end{equation}
 along with the other equations in equation(\ref{eq:new-set2}). Following similar methods as mentioned above, we solve the equations numerically, with suitable initial conditions at the epoch of nucleosynthesis. 
 %With the intention to follow standard cosmological evolution for matter and radiation, 
 We set the initial conditions similar to the previous case ($x=0.003,~y=1.5\times 10^{-14},~\lambda =1 ~{\text{and}}~r=0.0.99997$ at $\ln(1+z)=18.8$. Note that negative initial values for $\lambda$ are not permissible in this case due to the form of potential.) We did a similar analysis of the cosmological parameters as we have done in the case of $n=2$. Though the form of the potential of $V(\phi)$ in this case is different from the Starobinsky model, interestingly, we find that the behavior of the density parameters, the equation of state and the scalar field are similar to the Starobinsky model over the entire period of evolution (Fig \ref{fig:n-2}). The scalar field remains positive all along and the only difference is that it rolls down the potential $V(\phi)$. Further investigations on the modified gravity potential $V(\phi)$ and coupling potential $V_m$ reveal $V_m$ dominates over $V(\phi)$ like the previous case. Hence the scalar field rolls down the effective potential $V_{eff} (=V(\tilde\phi)+V_m)$ which is $\approx V_m$ (Fig \ref{fig:n-2p}). So, even though the form of the potential of $V(\phi)$ is different from the Starobinsky model,  the cosmological parameters behave similarly, and the scalar field remains positive. For all other values of n, ($\neq 2$), we get the same results.\\
 
{\bf \subsection{ $f (R) = R +\alpha_{n} R^n$ models with Cosmological Constant}}

We have also carried out the entire study including the Cosmological Constant $\Lambda$ along with the other matter components in the observable universe i.e, $\tilde\rho_M=\tilde\rho_r+\tilde\rho_m+\tilde\rho_{\Lambda}$.
%Following (\ref{eq:sf3}), 
The scalar field $\tilde{\phi}$ also couples with $\Lambda$ also, apart from matter. So, equations (\ref{eq:h2n}) and (\ref{eq:phi2n}) changes to include $\Lambda$, while  (\ref{eq:rhom2}) and (\ref{eq:rhor}) remain unchanged. Also, we have one more coupled equation for $\Lambda$.
\begin{equation}
\begin{aligned}
3\tilde H^{2} = \kappa^{2}\bigg[\frac{1}{2}{\tilde\phi^{\prime 2}} \tilde H^{2}+ V(\tilde\phi) + \tilde{\rho}_m+\tilde\rho_r +\tilde\rho_\Lambda\bigg]
\end{aligned}
\label{eq:h2nl}
\end{equation}
\begin{equation}
\begin{aligned}
\tilde\rho_\phi^\prime+ 3 (\tilde \rho_{\tilde\phi} + \tilde P_{\tilde\phi})  = \frac{\kappa}{\sqrt{6}} (\tilde{\rho}_m+4 \tilde{\rho_\Lambda})\tilde\phi^\prime
\end{aligned}
\label{eq:phi2nl}  
\end{equation}
\begin{equation}
\begin{aligned}
\tilde\rho_\Lambda^\prime = -4\frac{\kappa}{\sqrt{6}}\tilde{\rho}_\Lambda \tilde\phi^\prime
\end{aligned}
\label{eq:rhol}
\end{equation}
To proceed with the analysis, we define one more dynamical variable $l = \frac{\kappa^2 \tilde \rho_\Lambda}{3 \tilde H^2}$. In terms of all the dynamical variables, the new set of equations to be dealt with looks like
\begin{equation}
\begin{aligned}
\frac{dx}{dN} &= \frac{x}{2}(3x^2+3-3y^2 +r - 3l)-3x+\sqrt{\frac{3}{2}} \lambda y^2\\
& +\frac{1}{2}(1-x^2-y^2-r-l)+2l\\
\frac{dy}{dN} &= -y[\sqrt{\frac{3}{2}} x\lambda +\frac{1}{2}(3y^2-3-3x^2-r+3l)]\\
%\frac{d\lambda}{dN} =\sqrt{\frac{2}{27}}x(\sqrt{6}-6\lambda)(\sqrt{8/3}-\lambda)
\frac{d\lambda}{dN} &=\sqrt{6}x\lambda^2[1-\Gamma]\\
\frac{dr}{dN} &=-r[4+(-3x^2+3y^2-r+3l-3)]\\
\frac{dl}{dN} &=-l[4x+(-3x^2+3y^2-r+3l-3)]
\end{aligned}    
\label{eq:new-set3}
\end{equation}
We perform a similar numerical analysis with the new set of equations for the potential given by (\ref{eq:v}) with different $n$ as has been done without $\Lambda$. %with the suitable initial conditions so as to maintain the period of matter-radiation equality and matter energy density permitted by the standard Cosmology.
 Surprisingly, we found that inclusion of the Cosmological constant does not effect the evolution of different components. 
 %Now we have two terms in the right hand side of (\ref{eq:phi2nl}), which together contribute to the effective potential along with $V(\tilde\phi)$.
 Due to the inclusion of $\Lambda$, the scalar field couples with both matter and $\Lambda$ (equation (\ref{eq:phi2nl})). So we have two coupling terms $V_m$ and $V_l$.  But, even then, like the earlier scenario, $V_m$ is more dominant than both $V_l$ and $V(\tilde\phi)$ of modified gravity.
 %In fact in the figure, $V(\tilde\phi)$ and $V_l$ are superposed as both are  very close to 0. 
 Hence driven mostly by the matter coupling term $V_m$, the evolution of different components remain similar. The value of the scalar field for different $n$ also matches our earlier findings. For both $n=2$ and $n\neq2$, the scalar field remains positive throughout even with the inclusion of $\Lambda$ as well. Here we present only one case $n=2$ as an example.\\
 
\begin{figure}[h!]
\centering
\includegraphics[height=4.0cm, width=0.34\linewidth]{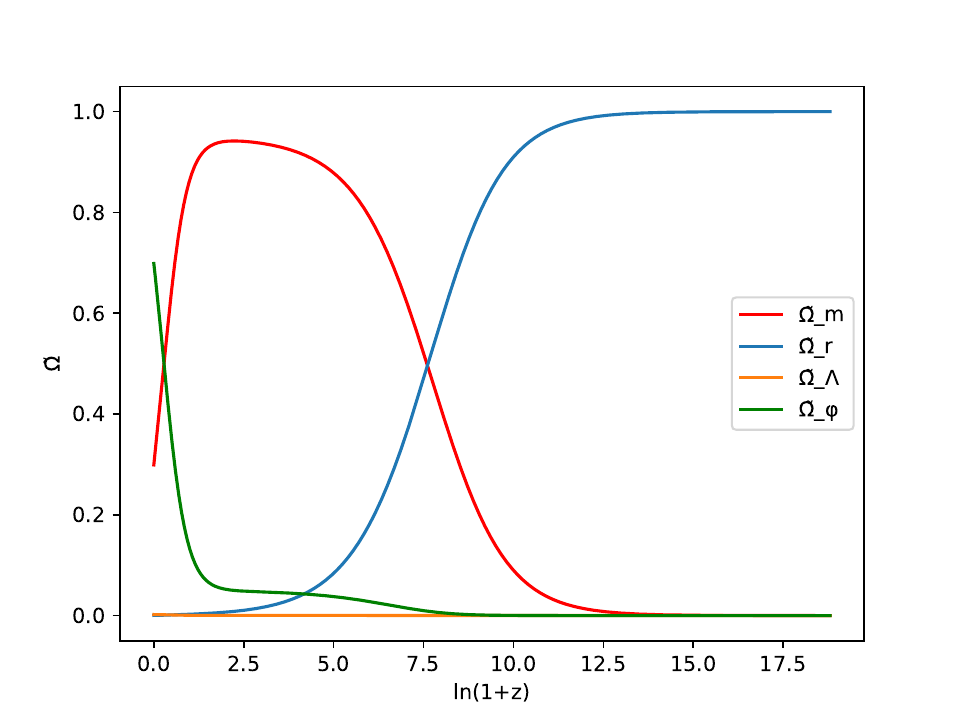}
\includegraphics[height=4.0cm, width=0.32\linewidth]{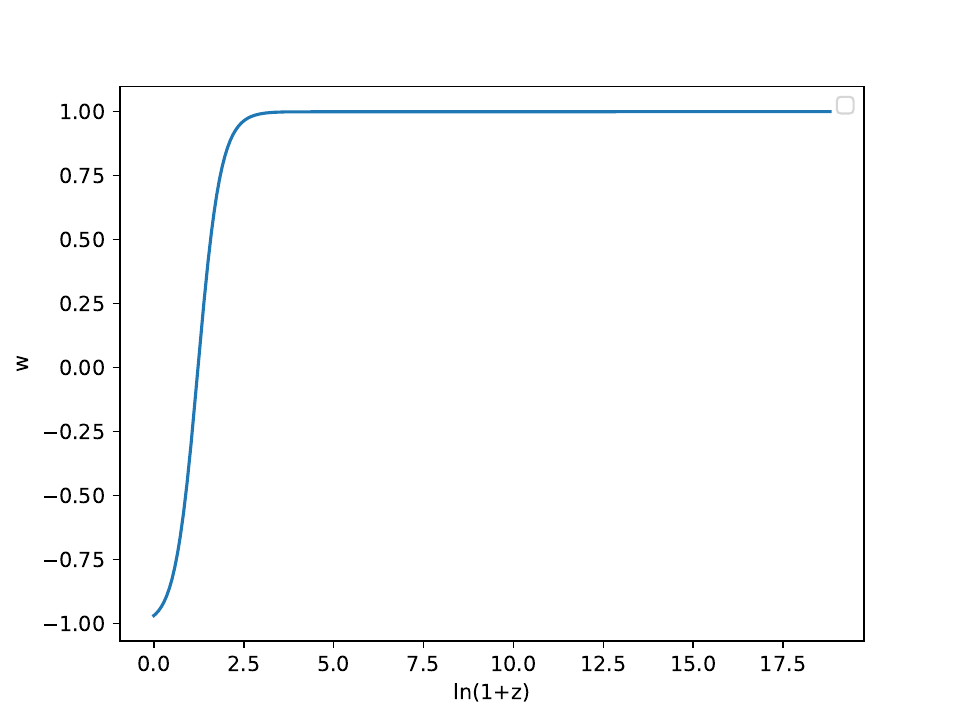}
\includegraphics[height=4.0cm, width=0.32\linewidth]{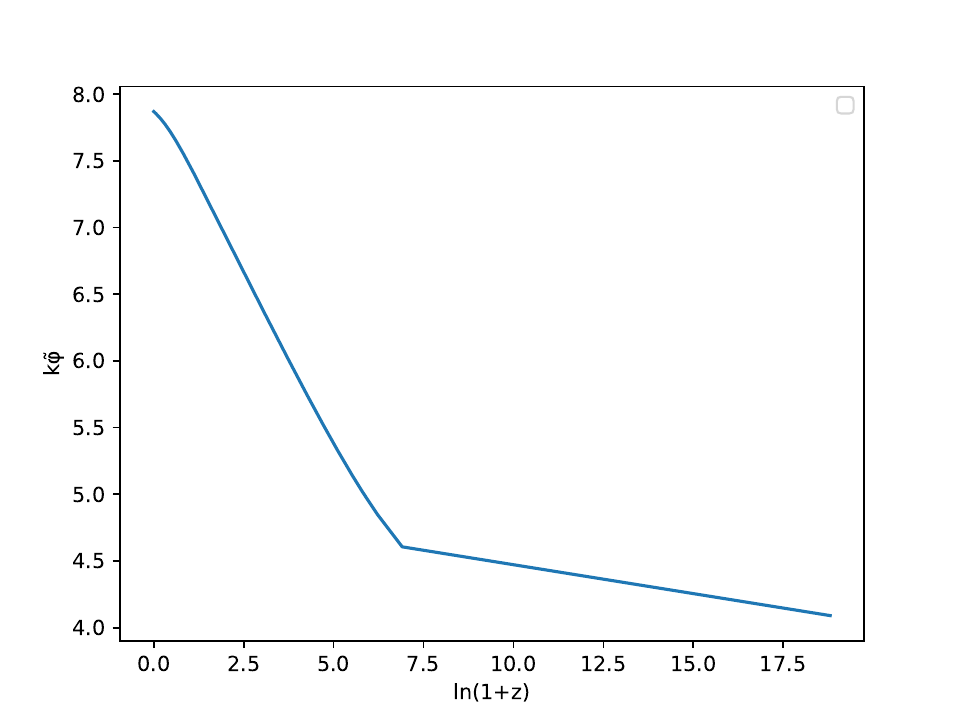}

\includegraphics[height=4.0cm, width=0.34\linewidth]{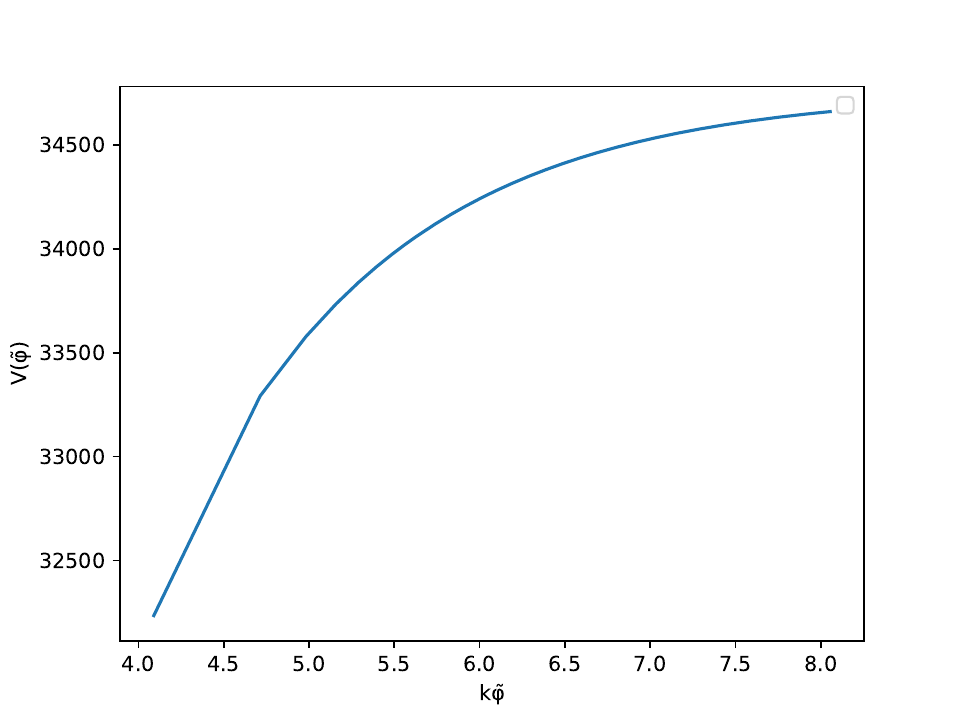}
\includegraphics[height=4.0cm, width=0.32\linewidth]{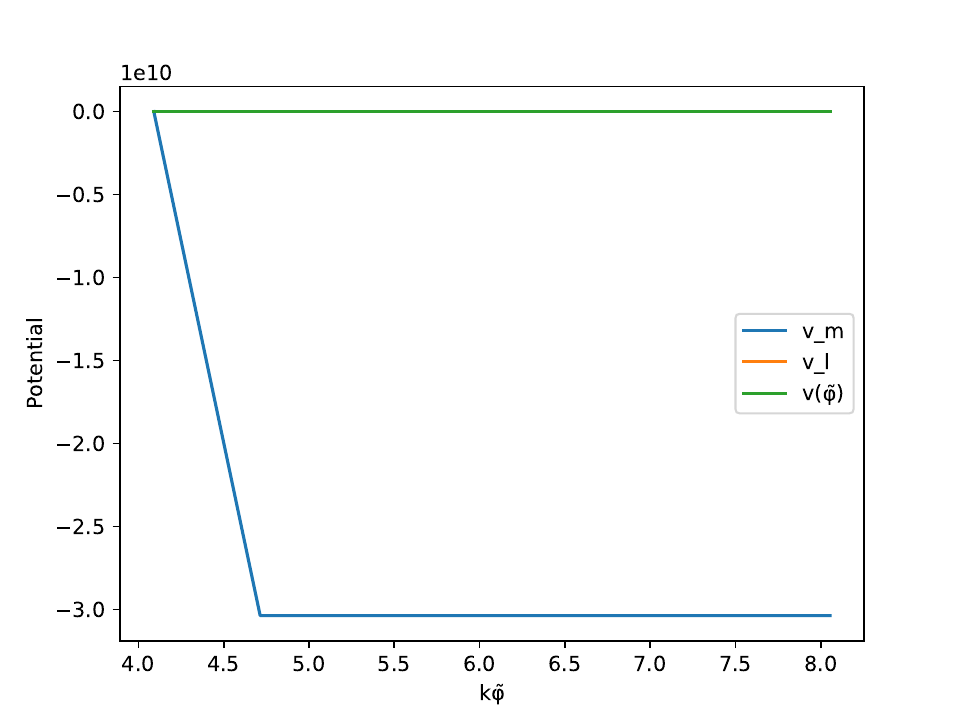}
\includegraphics[height=4.0cm, width=0.32\linewidth]{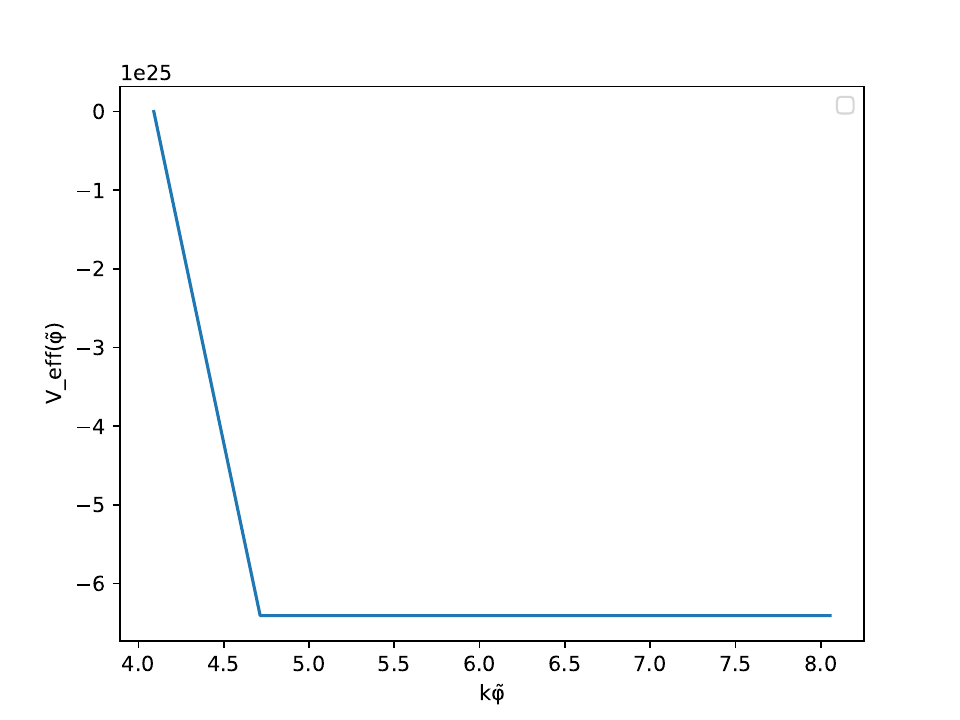}
 \caption{Evolution of different cosmological parameters and potentials including $\Lambda$ for $n=2$. The initial conditions in these plots are same as mentioned in section 5.1 along with $l=6\times 10^{-29}$ at $\ln(1+z)=18.8$ for the new variable $l$.}
\label{fig:lam2}
\end{figure}

\section{Results}
We have considered a very general class of modified gravity of the form $f(R)=R+\alpha_n R^n$ in an early universe around the epoch of nucleosynthesis. When conformally transformed in Eintein frame, modified gravity is equivalent to Einstein gravity along with a scalar degree of freedom $\tilde\phi$ following the potential given by (\ref{eq:v}). Depending on the behaviour of the potential we divided our cases in two categories $n=2$ (Case I) and $n\neq2$ (Case II). We investigate the evolution of different components like scalar field, radiation and matter in a background isotropic and homogeneous universe represented by FRW metric. After the whole analysis we found that both in the cases of $n=2$ and $n\neq2$, $\tilde\phi$ remains positive along the entire evolution and does not vary much. The matter - scalar field coupling forces the scalar field to roll down an effective potential. Such values of $\tilde\phi$ for the last two terms in action (\ref{eq:action4}) provide an additional exponential suppression. This makes the signature of the antisymmetric fields (both rank 2 and 3) heavily suppressed thereby failing to leave some imprints in the observable universe. In case of Starobinsky model, the scalar field attains higher positive value compared to the other cases, which means heavier suppression in the Starobinsky model.
We also performed the analysis adding the Cosmological constant $\Lambda$ along with the earlier components in the FRW background. Although the equations change significantly, the nature of the scalar field for different $n$ matches our earlier findings. 

 Thus, with or without $\Lambda$, the antisymmetric fields of rank 2 and 3 get heavily suppressed in the modified gravity models $R +\alpha_{n} R^n$, justifying the invisibility of these fields. Starobinsky model provides higher suppression compared to the other models.

\section{Discussion and Conclusion}
In our current universe, we don't see any signs of higher rank antisymmetric tensor fields having an impact on natural events. 
%though theoretically, we find that the coupling of KR field (2nd rank antisymmetric fied) is of the order of 1/$M_p$, equivalent to the coupling of gravity. 
We explain the suppression of these fields in the light of the higher curvature $f(R)$ theory. It has been extensively discussed in literature that modified $f(R)$ gravity when expressed in the Einstein frame, is equivalent to Einstein gravity along with a scalar degree of freedom $(\tilde\phi)$ which is the manifestation of higher curvature. The scalar field $(\tilde\phi)$ follows a potential governed by the form of $f(R)$.  Thus, in the Einstein frame, $\tilde\phi$ couples with all the other fields through the conformal transformation of the metric. 
%So, the field equations suitably change to accommodate the conservation of the energy momentum tensors.
In this work we study from an early universe post reheating epoch, how different components of energy evolve in a background isotropic and homogeneous universe represented by FRW metric. Further, we explore how the nature of the scalar field influences the coupling of the massless modes of higher rank antisymmetric fields present in the action (\ref{eq:action4}). The coupling, in turn, is responsible for the observation of the possible imprints of massless modes of these  antisymmetric fields.
  
For this study, we have considered the form of $f(R)$ to be $R +\alpha_{n} R^n$, where $\alpha_n$ has suitable dimensions.  The potentials $V(\tilde{\phi})$ generated from this form of modified gravity for different values of $n$ is shown in fig(\ref{fig:potn}). The potentials have minima at $\tilde{\phi}=0$ irrespective of the value of $n$. For $n=2$, the Starobinsky model, the potential increases and saturates to a constant for increasing positive $\tilde{\phi}$. With $\tilde{\phi}<0$ the potential increases exponentially. For $n\neq2$, the form of the potentials $V(\tilde\phi)$ are similar to each other. With increasing positive $\tilde{\phi}$ these potentials increase, reach a maxima, then decrease and become 0. For negative $\tilde{\phi}$ the potentials are not well defined, so we have plotted the absolute value of the potentials for $\tilde{\phi}<0$. The scalar field $\tilde{\phi}$ with these variety of potentials couples with different form of matter in an isotropic and homogeneous background and follow coupled non-linear field equations. 
%the form of the potential $V(\tilde\phi)$ remain same for all values of $n\neq2$ - the potential is zero for positive $\tilde\phi$ and increases exponentially for $\tilde\phi<0$. But, the $n=2$  Starobinsky model is different from others, as the potential is positive for positive $\tilde\phi$ and negative for negative $\tilde\phi$ with a minima at $\tilde\phi=0$. The field equations with different kinds of matter and scalar field in the homogeneous and isotropic background are coupled nonlinear differential equations.
We rewrite these field equations as a set of autonomous system of first order differential equations to tackle the complexity. The autonomous system is solved numerically, with suitable initial conditions at early epoch of nucleosynthesis. 
%such that, following standard model of cosmology, the matter radiation equality occurs around redshift $z\sim 3300$ and the matter content  is $30\%$ of the total in the present universe.

Though we studied different cases for $-4\geq n\geq 4$,  we broadly divide them into two categories $n=2$ and $n\neq 2$, depending on the behaviour of the potential. We found that the evolution of all components mimic their behaviour in standard model of cosmology and the scalar field $\tilde\phi$ is subdominant throughout the evolution of the universe until recently. Further investigation of the equation of state $\omega_{\tilde\phi}$ reveals in the late time $\tilde\phi$ behaves as a source of dark energy, which resemble the behaviour of thawing quintessence model. Surprisingly we obtain similar behaviours for all of these parameters for both the categories $(n=2~{\text{and}}~n\neq 2)$, even though the potentials are different. The reason for such similarity is that the dynamics is dominated by the coupling between matter and $\tilde\phi$, which gives rise to an effective potential. The effective potential exhibit similar nature for different values of $n$ and the variation of potential $V(\tilde\phi)$ does not have any significant impact on the evolution of the parameters.

 Interestingly, for different values of $n$  (both  $n=2$ and $n\neq 2$ ), the scalar field evolves in a similar fashion, and remain positive throughout. This happens so because the $\tilde\phi$ rolls down the effective potential.  
%The value of $\tilde\phi$ depends on the slope of the potential. As discussed earlier, equation (\ref{eq:phi1}) reveals that the scalar field $\tilde\phi$ rolls down the potential but does not reach the minima due to the coupling with matter which is dominating. 
For $n=2$, $\tilde\phi$ climbs up $V(\tilde{\phi})$ due to the coupling with matter and is non-zero at present
where as, for $n\neq 2$, $\tilde\phi$ rolls down  $V(\tilde{\phi})$ due to the coupling and is zero at present.  However, for both the cases the value of $\tilde\phi$ is positive and it increases with redshift. 
%Again this time also, it cannot reach zero due to the coupling. 
Hence not only both the categories agree with standard model of cosmology, for both of them $(n=2~{\text{and}}~n\neq2)$, the last two terms of the action (\ref{eq:action4}) get an extra exponential suppression. 
%in addition to the usual $1/M_P$ coupling. For $n\neq2$, we obtain the enhancement of the coupling term instead of the suppression. 
As there has been no experimental evidence of the footprint of antisymmetric fields on the present Universe,
%so $n=2$ i.e., the Starobinsky model is the only model in a 
the general class of modified gravity where $f(R)=R +\alpha_{n} R^n$ provides an explanation of the suppression of such fields. In case of $n=2$ the Starobinsky model, the scalar field attains higher positive value compared to the other cases, which means the Starobinsky model supports heavier suppression.
We have also added cosmological constant $\Lambda$ as another component. Even though $\Lambda$ also couples with $\tilde\phi$, the effect is very less with respect to matter-$\tilde\phi$ coupling which remains the dominant term in the whole analysis. Hence, the dynamics and inference regarding the suppression of the antisymmetric fields remain unchanged. So with or without $\Lambda$ the absence of the signatures of the massless modes of antisymmetric fields can be justified by a general class of modified $f(R)$ gravity , with Starobinsky model predicting the heavier suppression. 

\section{Acknowledgement}

Sonej Alam acknowledges the CSIR fellowship provided by the Govt. of India under the CSIR-JRF scheme (file no. 09/0466(12904)/2021). The authors thank Wali Hossain and Anjan A. Sen for their invaluable insights and suggestions.

\bibliographystyle{JHEP} % Specify your bibliography style
\bibliography{main}

\end{document}